\documentclass[pre,twocolumn,groupedaddress]{revtex4-1}
\usepackage[utf8]{inputenc}
\usepackage{amsmath}
\usepackage{amsfonts}
\usepackage{amssymb}
\usepackage{graphicx}
\usepackage{graphics}
\usepackage{color}

\begin{document}
\title{Optimized efficiency at maximum $\dot{\Omega}$ figure of merit and efficient power of \\
Power law dissipative Carnot like heat engines} 
\author{K. Nilavarasi$^1$  and  M. Ponmurugan$^2$}
\email[]{ponphy@cutn.ac.in ;\ nilavarasikv@gmail.com}
\affiliation{$^1$Department of Physics, National Institute of Technology-Thiruchirapalli,
Thiruchirapalli- 620 015, Tamil Nadu, India. \\
$^2$Department of Physics, School of Basic and Applied Sciences, Central
University of Tamil Nadu, Thiruvarur 610 005, Tamil Nadu, India.}
\date{\today}
\begin{abstract}

In the present work, a power law dissipative Carnot like heat engine cycle of two irreversible isothermal and two irreversible adiabatic processes with finite time non-adiabatic dissipation is considered and the efficiency under two optimization criteria $\dot{\Omega}$ figure of merit and efficient power, $\chi_{ep}$ is studied. The generalized extreme bounds of the optimized efficiency under the above said optimization criteria are obtained.
The lower and upper bounds of the efficiency for the low dissipation Carnot-like heat engine  under these optimization criteria are obtained with dissipation level $\delta$ = 1. In corroborate with efficiency at maximum power, this result also shows the presence of non-adiabatic dissipation does not influence the extreme bounds on the efficiency optimized  by both these target functions in the low dissipation model.

\end{abstract}

\maketitle

\section{Introduction}
The world wide problem concerning the energy and dwindling fossil fuels created a renewed interest in optimizing the efficiency of heat engines. Since heat engines, the crucial components of the Industrial revolution, convert heat energy to mechanical energy.  Thus finding the more realistic upper bounds on the efficiency would pave the way for the reduction of the energy consumption in heat
engines and hence a suitable solution to the concerns related to the existing energy problem. Carnot heat engine, an ideal one has the maximum efficiency, $\eta_{C} = 1-(\frac{T_{c}}{T_{h}})$, where, $T_{c}$ and $T_{h}$ are the temperatures of cold and hot reservoir, respectively. The Carnot cycle consumes infinite time to complete the process and hence its power output is zero. 

The real heat engines operate in finite time duration with non-zero power output whose efficiency is always bounded below the ideal Carnot efficiency. In order to obtain a non zero  power, the 
thermodynamic processes for a heat engine should  take place in a finite time duration with optimized efficiency. Finite time thermodynamics (FTT) is one such wider field of thermodynamic optimization providing more realistic bounds on the efficiency of real systems by considering finite time irreversible processes \cite{andersen1, anderson2}. The theoretical bounds determined from the FTT provide the optimal conditions for designing the real systems.  Yvon \cite{yvon}, Novikov \cite{novikov}, Chambadal \cite{Chambadal} and later Curzon and Ahlborn \cite{curzon} are the pioneers in obtaining the efficiency optimized at maximum power by using 
endo-reversible concept in the reversible Carnot cycle. The so-called Curzon-Ahlborn expression for the efficiency at maximum power obtained from the above model is given by, $\eta_{CA} = 1-(\frac{T_{c}}{T_{h}})^{1/2}$ \cite{curzon}. 

Recently, there has been tremendous progress in identifying the performance limits of thermodynamic processes through various optimization of thermodynamic cycles of heat engine using finite-time thermodynamics \cite{van, salas, zhangguo, medina}. In particular, irrespective of any heat transfer laws, Esposito et.al. obtained the extreme bounds on the efficiency at maximum power for a low dissipation Carnot like heat engines \cite{esposito}. The main assumption of their model is that the irreversible entropy production in each isothermal process is inversely proportional to the time required for completing that process.  On the other hand, Ma used the parameter called per-unit time efficiency optimized at maximum power and  obtained the extreme bounds on the efficiency \cite{ma}. This criterion was found to be a compromise between the efficiency and speed of the thermodynamic cycle.  

Few heat engine studies based on the low dissipation reported the extreme bounds on the efficiency of Carnot like heat engines with the consideration of non-adiabatic dissipation in finite time adiabatic
 processes \cite{he1,he2}. The dissipation that occurs due to the effects of inner friction during the finite time adiabatic process is known as non-adiabatic dissipation \cite{he1,infriction,refrig}.
These investigations showed that the additionally incorporated non-adiabatic dissipation does not influence the extreme bounds on the efficiency at maximum power. Followed by the investigation of W. Yang and Z. Tu \cite{chun}, one of the present author, obtained the generalized bounds on the efficiency at maximum power by incorporating the power law dissipation in finite time Carnot like heat engine model \cite{pon2} and also showed that the generalized extreme bounds are not influenced by the additionally incorporated non-adiabatic dissipation \cite{pon}. In all these studies, optimization of efficiency at maximum power is used to investigate the performance of heat engines. 

 Even though, the efficiency at maximum power is a desirable operational regime, several other optimization parameters are also used to enhance the heat engine performance. In particular, $\dot{\Omega}$ figure of merit and efficient power, $\chi_{ep}$ are two such criteria that has attracted much attention nowadays to study the heat engine performance.  The former, $\dot{\Omega}$ figure of merit is defined as
\cite{rocco} $\dot{\Omega}=(2\eta-\eta_{max})\dot{Q_{h}}$, where $\eta$ is the efficiency, $\eta_{max}$ is the maximum efficiency and $\dot{Q_{h}}$ is the rate of heat flow or heat exchanged with the hot reservoir per cycle time \cite{sancheez}. The $\dot{\Omega}$ figure of merit unifies the trade-off between the useful energy delivered and energy lost of heat engines \cite{rocco}.
Whereas the latter, efficient power provides a compromise between efficiency 
and power $P$, which is defined as $\chi_{ep} = \eta P$ \cite{stucki,yilmaz,holubec}.

 The extreme bounds on the efficiency by optimizing both these target functions 
were studied for low dissipation Carnot like heat engines without incorporating non-adiabatic dissipation \cite{lowOmega,singh}.
This raises a question whether the inclusion of non-adiabatic dissipation will influence the extreme bounds on the efficiency
in the low dissipation model by optimizing both these target functions. Although the low-dissipation model is a well-founded model for many heat engines \cite{holubec, schmiedl}, it has been observed that this model might not be suitable for real heat engines operating at different dissipation levels \cite{chun, pon2, holubec, medinasl}.  Hence the present work investigates the generalized  minimum  and maximum  bounds on the efficiency at maximum $\dot{\Omega}$ figure of merit and efficient power, $\chi_{ep}$ of power law dissipative Carnot like heat engines \cite{pon2} which incorporates the generalized dissipation and also addresses the influence of non-adiabatic dissipation on these extreme bounds in the low dissipation model as a special case.
  
This paper is organized as follows: In section II, the model of power law dissipative Carnot like heat engine is explained. In section III and IV, the optimization of efficiency at maximum $\dot{\Omega}$ figure of merit and at maximum efficient power are derived and its extreme bounds are discussed. The paper concludes with the conclusion in section V.

\begin{figure}
\includegraphics[width=3.75in,height=3.75in]{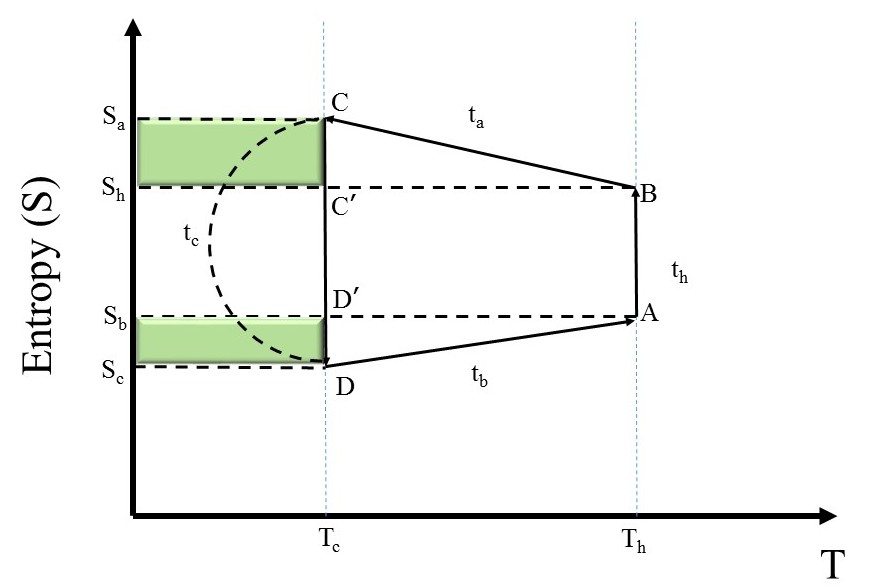}
\caption{
Figure represents the temperature $(T)$ and entropy $(S)$ 
plane of an irreversible Carnot-like cycle. The system is in contact with the hot (cold) reservoir during the time interval $t_h$ ($t_c$) and 
$t_a$ ($t_b$) represents the time duration for the adiabatic expansion 
(compression). The value of  entropy at the time the system completes a particular process is denoted by $S_i$ ($i:a,b,c,h$).
Work done by the system to overcome the non-adiabatic dissipation 
in the finite time adiabatic expansion (top) and compression (bottom) processes are  shown in the shaded (rectangles) areas \cite{he1,he2}.
}\label{tsfig}
\end{figure}

\section{Power law dissipative Carnot like heat engine}

Figure \ref{tsfig} represents the temperature $(T)$ and entropy $(S)$ 
plane of an irreversible Carnot-like cycle ($A \to  B \to
C \to D \to A$) \cite{he2}. The system is in contact with the hot (cold)
reservoir at the constant temperature $T_h$ ($T_c$) in the finite time interval $t_h$ ($t_c$) during $A \rightarrow B$ ($C \rightarrow D$)  for an isothermal expansion 
(compression) processes. During $B \to C$ ($D \to A$) in which the system is decoupled from the hot (cold) reservoir and undergoes a finite time adiabatic expansion (compression) process in a finite time duration $t_a$ ($t_b$). The value of entropy at the time the system completes a particular process is denoted by $S_i$ ($i:a,b,c,h$). Here ($A \to  B \to
C' \to D' \to A$) represents the (reversible) Carnot cycle in which 
$S_a = S_h$ and $S_b = S_c$.

In this work, a power law dissipative Carnot like heat engine cycle of two irreversible isothermal and two irreversible adiabatic processes with finite time non-adiabatic dissipation is considered \cite{pon2,pon}.
The four processes involved in the present model are discussed below: 
\begin{itemize}
	\item Isothermal expansion:
During this process, the working substance is in contact with the hot reservoir at a higher temperature $T_{h}$ for a time interval $t_{h}$.
In this process there is an exchange of $Q_{h}$ amount of heat between the working substance and the hot reservoir and the change in entropy is given as
\begin{equation}
\Delta S=\Delta S_h=Q_{h}/T_h+\Delta S^{ir}_{h}
\label{eq1:}
\end{equation}
where $\Delta S^{ir}_{h}$ being the irreversible entropy production during isothermal expansion process.
\item Adiabatic Expansion:
The non-adiabatic dissipation increases the entropy during this adiabatic expansion process and the irreversible entropy production during the time interval $t_{h} < t < t_{h} + t_{a}$ is denoted by \cite{he1},
\begin{equation}
\Delta S^{ir}_{a}=S_{a} - S_{h}
\label{eq2:}
\end{equation}
where $S_{a}$ and $S_{h}$ denotes the entropy at the instant $t_{a}$ and $t_{h}$, respectively.
\item Isothermal Compression:
Now the the working substance is in contact with the low temperature $(T_{c})$ cold reservoir for the time period $t_{h} + t_{a} < t <
t_{h} + t_{a} + t_{c}$ with the exchange of $Q_{c}$ amount of heat between the working substance and the cold reservoir. The change in  entropy is given as, 
\begin{equation}
\Delta S_c = Q_{c}/T_c +\Delta S^{ir}_{c}
\label{eq3:}
\end{equation}
where $\Delta S^{ir}_{c}$ being the irreversible entropy production during isothermal compression process.
\item Adiabatic Compression: 
In the process of adiabatic compression during the time interval $t_{h} +t_{a} +t_{c} < t < t_{h}+t_{a}+t_{c}+t_{b}$, the working substance is removed from the cold reservoir and now the entropy production due to non-adiabatic dissipation is given by \cite{he1},
\begin{equation}
\Delta S^{ir}_{b}=S_{b} - S_{c}
\label{eq4:}
\end{equation}
where $S_{b}$ and $S_{c}$ denotes the entropy at the instant $t_{b}$ and $t_{c}$, respectively.
\end{itemize}

At the instance of completing the single cycle, the
system recovers to its initial state and the total
change in entropy of the system is zero \cite{esposito,he1,he2}, i.e.,
$\Delta S + \Delta S^{ir}_{a} + \Delta S^{ir}_{b} + \Delta S_c =0$. 
Therefore
$\Delta S_c= -(\Delta S + \Delta S^{ir}_{a} + \Delta S^{ir}_{b})$.
Since the present model also considers the finite time non-adiabatic process, there will be an additional irreversible entropy production $\Delta S^{ir}_{a}$ and  $\Delta S^{ir}_{b}$ during the adiabatic 
processes \cite{he1,he2}. From Eq.(\ref{eq1:}) and Eq.(\ref{eq3:}), the amount of heat $Q_h$ and $Q_c$ exchanged between the hot and cold reservoirs and the working substance are obtained as \cite{he1,he2}
\begin{equation}
 Q_{h}=T_h \{ \Delta S-\Delta S^{ir}_{h}\}
\label{eqh:}
\end{equation}
\begin{equation}
 Q_{c}=T_c \{-\Delta S-\Delta S^{ir}_{c}-\Delta S^{ir}_{a}-\Delta S^{ir}_{b}\}.
\label{eqc:}
\end{equation}

Even though many studies incorporated the $1/\tau$ scaling of the
irreversible entropy production both in a finite-time isothermal
process \cite{esposito,scal1} and  a finite-time adiabatic process \cite{he1,he2,refrig},
a recent theoretical study on quantum Otto engine showed (in terms of extra adiabatic
work) $1/\tau^2$ scaling of the irreversible entropy production in a
finite-time adiabatic process \cite{scal2}. Here $\tau$ is the 
controlling or contact time in which each processes take place. 
This raises a possibility that the irreversible entropy production may  
scales with $\tau$ with other values of exponent for various real heat engines. 
Considering the above facts, the more generalized  
power law dissipative Carnot like heat engines has been proposed earlier 
and studied in detail for efficiency at maximum power \cite{chun,pon2,pon}.

The irreversible entropy production associated with the isothermal processes 
and the adiabatic processes can be written in a generalized power law 
dissipative form as \cite{pon}, 
\begin{equation}
\Delta S^{ir}_{i} =\alpha_{i}\left(\frac{\sigma_{i}}{t_{i}}\right)^{1/\delta},
\label{deltas1}
\end{equation}
where $i:h,a,b,c$ and $\sigma_{i}=\lambda_{i}\Sigma_{i}$, in which $\alpha_{i}$ and $\lambda_{i}$ are the tuning parameters and $\Sigma_{i}$ are the isothermal and adiabatic dissipation coefficients \cite{he1,he2}.  The level of dissipation present in the system is signified by the value of $\delta$ \cite{pon2} in which $\delta =1$ represents normal or low-dissipation regime,
$0 < \delta < 1$: Sub dissipation regime and  $\delta > 1$ : Super dissipation regime \cite{chun}. 
It should be noted that the employed model contains parameter $\alpha_i$, which might be related to the control scheme that tune the system energy levels during the isothermal and adiabatic processes \cite{modylow1}
and the parameter $\lambda_i$ are related with some external controlled parameter that drives the system during the isothermal and adiabatic processes in a given time interval \cite{chun,modylow2}. A suitable combination of control schemes \cite{modylow1,modylow2}
can be employed to control the irreversible entropy generation by using these tuning parameters.

Thus, the expression for amount of heat exchanged $Q_{h}$ and $Q_{c}$ can be rewritten as \cite{he1,pon},
\begin{equation}
Q_{h}=T_{h}\left\{\Delta S-\alpha_{h}\left(\frac{\sigma_{h}}{t_{h}}\right)^{1/\delta}\right\}
\label{qh}
\end{equation}
and 
\begin{equation}
Q_{c}=T_{c}\left\{-\Delta S-\sum_{i=a,b,c}\alpha_{i}\left(\frac{\sigma_{i}}{t_{i}}\right)^{1/\delta}\right\}.
\label{qc}
\end{equation}

During the total time period $t = t_{h} +t_{c} +t_{a} +t_{b}$, the work performed by the engine is given by, $-W = Q_{h}+Q_{c}$. Throughout this paper, the convention used is that the work done and heat absorbed by the system are positive. The power generated during the Carnot cycle
is $P=\frac{-W}{t}$.  On substituting the values of $Q_{h}$ and $Q_{c}$, the expression for power can be written as,
\begin{widetext}
\begin{equation}
P = \frac{1}{t}\left\{(T_{h}-T_{c})\Delta S -T_{h}\alpha_{h}\left(\frac{\sigma_{h}}{t_{h}}\right)^{1/\delta}-T_{c}\sum_{i=a,b,c}\alpha_{i}\left(\frac{\sigma_{i}}{t_{i}}\right)^{1/\delta}\right\}.
\label{power}
\end{equation}
The efficiency of the heat engine is then given by, 
\begin{equation}
\eta = \frac{Q_{h}+Q_{c}}{Q_{h}}
\label{etainital}
\end{equation}
which on substituting the values of $Q_{h}$ and $Q_{c}$ becomes \cite{pon},
\begin{equation}
\eta = \eta_{C}-\frac{T_{c}}{T_{h}\left[\frac{\Delta S}{\alpha_{h}}\left(\frac{t_{h}}{\sigma_{h}}\right)^{1/\delta} -1\right]} \left[1+\sum_{i=a,b,c}\left(\frac{\alpha_{i}}{\alpha_{h}}\right)\left(\frac{\sigma_{i}t_{h}}{\sigma_{h}t_{i}}\right)^{1/\delta}\right].
\label{eta}
\end{equation}
\end{widetext}
In the following sections, the $\dot{\Omega}$ figure of merit and the efficient power $\chi_{ep}$ are used as target functions and are optimized for analyzing the performance of heat engines with (isothermal and non-adiabatic) power law dissipation. 

\section{Efficiency at maximum $\dot{\Omega}$ figure of merit}
The $\dot{\Omega}$ figure of merit is a  trade-off function that provides a compromise between the
useful energy and the lost energy. It can be defined as $\dot{\Omega} = (2\eta -\eta_{max})P/\eta$, where $\eta_{max}$ is the maximum
efficiency of a heat engine which is nothing but the Carnot efficiency $\eta_{C}$ \cite{sancheez}. With the inclusion of $P=\frac{Q_{h}+Q_{c}}{t}$ and $\eta =\frac{Q_{h}+Q_{c}}{Q_{h}}$, one can obtain,
\begin{widetext}
\begin{equation}
\dot{\Omega} = (2\eta -\eta_{C}) \frac{Q_{h}}{t}.
\label{omega}
\end{equation}
On substituting the values of $\eta$, $\eta_{C}$ and $Q_{h}$ in Eq.(\ref{omega}), the expression for $\dot{\Omega}$ figure of merit can be rewritten as, 
\begin{equation}
\dot{\Omega} = \frac{1}{t}\left\{(T_{h}-T_{c})\Delta S - (T_{h}+T_{c})\alpha_{h}\left(\frac{\sigma_{h}}{t_{h}}\right)^{1/\delta}-2T_{c}\sum_{i=a,b,c}\alpha_{i}\left(\frac{\sigma_{i}}{t_{i}}\right)^{1/\delta}\right\}.
\label{omega1}
\end{equation}

Optimizing the $\dot{\Omega}$ figure of merit with respect to time $t_{i} (i : h, c, a, b)$ gives the values of $\tilde{t}_{i}$ at which $\dot{\Omega}$ figure of merit is maximum. The values for $\tilde{t}_{i}(i : h, c, a, b)$ by considering $\frac{\partial \dot{\Omega}}{\partial t_{i}}=0$ are given below:
\begin{equation}
\tilde{t}_{h}=\left[\frac{(T_{h}+T_{c})\alpha_{h}\sigma_{h}^{1/\delta}}{(T_{h}-T_{c})\Delta S}\left\{\left(1+\frac{1}{\delta}\right)\left(1+\sum_{i=a,b,c}\left(\frac{2T_{c}\alpha_{i}\sigma_{i}^{1/\delta}}{(T_{h}+T_{c})\alpha_{h}\sigma_{h}^{1/\delta}}\right)^{\frac{\delta}{\delta+1}}\right)\right\}\right]^{\delta},
\label{th1}
\end{equation}
\begin{equation}
\tilde{t}_{a}=\left[\frac{2T_{c}\alpha_{a}\sigma_{a}^{1/\delta}}{(T_{h}-T_{c})\Delta S}\left\{\left(1+\frac{1}{\delta}\right)\left(1+\left(\frac{(T_{h}+T_{c})\alpha_{h}\sigma_{h}^{1/\delta}}{2T_{c}\alpha_{a}\sigma_{a}^{1/\delta}}\right)^{\frac{\delta}{\delta+1}}+\sum_{i=b,c}\left(\frac{\alpha_{i}\sigma_{i}^{1/\delta}}{\alpha_{a}\sigma_{a}^{1/\delta}}\right)^{\frac{\delta}{\delta+1}}\right\}\right)\right]^{\delta},
\label{ta}
\end{equation}

\begin{equation}
\tilde{t}_{b}=\left[\frac{2T_{c}\alpha_{b}\sigma_{b}^{1/\delta}}{(T_{h}-T_{c})\Delta S}\left\{\left(1+\frac{1}{\delta}\right)\left(1+\left(\frac{(T_{h}+T_{c})\alpha_{h}\sigma_{h}^{1/\delta}}{2T_{c}\alpha_{b}\sigma_{b}^{1/\delta}}\right)^{\frac{\delta}{\delta+1}}+\sum_{i=a,c}\left(\frac{\alpha_{i}\sigma_{i}^{1/\delta}}{\alpha_{b}\sigma_{b}^{1/\delta}}\right)^{\frac{\delta}{\delta+1}}\right\}\right)\right]^{\delta},
\label{tb}
\end{equation}
and 
\begin{equation}
\tilde{t}_{c}=\left[\frac{2T_{c}\alpha_{c}\sigma_{c}^{1/\delta}}{(T_{h}-T_{c})\Delta S}\left\{\left(1+\frac{1}{\delta}\right)\left(1+\left(\frac{(T_{h}+T_{c})\alpha_{h}\sigma_{h}^{1/\delta}}{2T_{c}\alpha_{c}\sigma_{c}^{1/\delta}}\right)^{\frac{\delta}{\delta+1}}+\sum_{i=a,b}\left(\frac{\alpha_{i}\sigma_{i}^{1/\delta}}{\alpha_{c}\sigma_{c}^{1/\delta}}\right)^{\frac{\delta}{\delta+1}}\right\}\right)\right]^{\delta}.
\label{tc}
\end{equation}
\end{widetext}

The ratios of $\tilde{t}_{h}$ and $\tilde{t}_{i} (i: a,b,c) $ can also be obtained from the
optimized $\dot{\Omega}$ figure of merit and are given below:
\begin{equation}
\left(\frac{\tilde{t}_{h}}{\tilde{t}_{i}}\right)^{\frac{1}{\delta}+1}=\frac{(T_{h}+T_{c})\alpha_{h}\sigma_{h}^{1/\delta}}{2T_{c}\alpha_{i}\sigma_{i}^{1/\delta}}
\label{thti}
\end{equation}
Similarly the ratios for $\frac{\tilde{t}_{j}}{\tilde{t}_{i}}$ with $i,j = a, b, h$ are also given by,
\begin{equation}
\left(\frac{\tilde{t}_{i}}{\tilde{t}_{j}}\right)^{\frac{1}{\delta}+1}=\frac{\alpha_{i}\sigma_{i}^{1/\delta}}{\alpha_{j}\sigma_{j}^{1/\delta}}.
\label{titj}
\end{equation}
Using Eq.(\ref{th1}) and Eq.(\ref{thti}) on Eq.(\ref{eta}) provides the efficiency at maximum $\dot{\Omega}$ figure of merit, $\eta_{\dot{\Omega}}$ and is given by, 
\begin{equation}
\eta_{\dot{\Omega}} = \eta_{C} - \frac{\frac{T_{c}}{T_{h}}\left(1+\frac{T_{h}+T_{c}}{2T_{c}}\zeta\right)}{\left(\frac{T_{h}+T_{c}}{T_{h}-T_{c}}\right)\left(\frac{1}{\delta}+1\right)(1+\zeta)-1}
\label{maxeta}
\end{equation}
where, $\zeta = \sum_{i=a,b,c}\left\{\left(\frac{2T_{c}}{T_{h}+T_{c}}\right)\frac{\alpha_{i}\sigma_{i}^{1/\delta}}{\alpha_{h}\sigma_{h}^{1/\delta}}\right\}^{\frac{\delta}{\delta+1}}$.
It is observed that when neglecting the adiabatic dissipation co-efficients, $\sigma_{a}=0$ and $\sigma_{b}=0$, the efficiency of heat engine (Eq.\ref{maxeta}) reduces to the one derived for Carnot like heat engines without adiabatic dissipation \cite{sancheez}.  

It can be observed from Eq.(\ref{maxeta}), that the value of efficiency at the maximum $\dot{\Omega}$ figure of merit depends on the ratio between values of $\sigma_{i}$'s and $\sigma_{h}$. The generalized extreme bounds of the efficiency at maximum $"\dot{\Omega}"$ figure of merit are obtained from Eq.(\ref{maxeta}) as,
\begin{widetext}
\begin{equation}
\eta_{C}\left(1-\frac{1}{2\left(1+\frac{1}{\delta}\right)}\right)\equiv\eta^{-}_{\dot{\Omega}}\leq\eta_{\dot{\Omega}}\leq\eta^{+}_{\dot{\Omega}}\equiv\eta_{C}\left[\frac{(1-\eta_{C})\left(1+\frac{1}{\delta}\right)+\frac{1}{\delta}}{(1-\eta_{C})\left(2+\frac{1}{\delta}\right)+\frac{1}{\delta}}\right].
\label{limit}
\end{equation}
\end{widetext}

These generalized  lower and upper bounds of the efficiency at maximum $\dot{\Omega}$ figure of merit are achieved when $\sigma_{h}\rightarrow 0$ and $\sigma_{h}\rightarrow \infty$. When $\delta= 1$, the lower bound becomes $3\eta_{C}/4$ for $\sigma_{h}\rightarrow 0$ and the upper bound becomes $\eta_{C}\left(\frac{3-2\eta_{C}}{4-3\eta_{C}}\right)$ for $\sigma_{h}\rightarrow \infty$, which is the bound of the efficiency of Carnot like heat engine at the maximum $\dot{\Omega}$ figure of merit obtained for the low dissipation case \cite{lowOmega}.
This shows that the inclusion of finite time non-adiabatic dissipation on low-dissipation model
does not influence the lower and upper bound on the efficiency optimized at maximum $\dot{\Omega}$ figure of merit.  
  It is to be noted that when $\delta \to 0$ (no dissipation), $\eta^{-}_{\dot{\Omega}}$ and $\eta^{+}_{\dot{\Omega}} \to \eta_C$ and when $\delta \to \infty$ (high super dissipation limit), $\eta^{-}_{\dot{\Omega}}$ and $\eta^{+}_{\dot{\Omega}} \to \eta_C/2$. This shows that $\dot{\Omega}$ figure of merit provides half the Carnot efficiency even at very high level (super) of  power law  dissipation which further confirms that $\dot{\Omega}$ figure of merit provides a compromise between the useful energy and the energy lost. Thus, a more generalized upper and lower bounds on the efficiency of Carnot like heat engine can be obtained under the combined adiabatic and isothermal power law dissipation in the asymmetric limits.

\section{Efficiency at maximum efficient power $\chi_{ep}$}
This section discusses the optimization of efficient power $\chi_{ep} = \eta P$ and its significance in detail. 
The efficient power  can be expressed using Eq.(\ref{etainital}) and the fact that $P=\frac{Q_{h}+Q_{h}}{t}$ as, 
\begin{equation}
\chi_{ep}=\eta P = \frac{(Q_{h}+Q_{c})^{2}}{Q_{h}t}
\label{etap}
\end{equation}
Using Eq.(\ref{qh}) and Eq.(\ref{qc}), the following relation for $\chi_{ep}$ is obtained.  
\begin{widetext}
\begin{equation}
\chi_{ep}= \frac{\left[(T_{h}-T_{c})\Delta S -T_{h}\alpha_{h}\left(\frac{\sigma_{h}}{t_{h}}\right)^{1/\delta}-T_{c}\sum_{i=a,b,c}\alpha_{i}\left(\frac{\sigma_{i}}{t_{i}}\right)^{1/\delta}\right]^{2}}{tT_{h}\left(\Delta S-\alpha_{h}\left(\frac{\sigma_{h}}{t_{h}}\right)^{1/\delta}\right)}.
\label{etap2}
\end{equation}

Similar to the $\dot{\Omega}$ figure of merit, the optimizing the efficient power $\chi_{ep}$ with respect to the time $t_{i}(i : h, c, a, b)$ gives the values of $\tilde{t}_{i}$ at which $\chi_{ep}$ is maximum. The values for $\tilde{t}_{i}(i : h, c, a, b)$ by considering $\frac{\partial \chi_{ep}}{\partial t_{i}}=0$  are given below:
\begin{equation}
\tilde{t}_{h}=\left[\frac{T_{h}\alpha_{h}\sigma_{h}^{1/\delta}}{(T_{h}-T_{c})\Delta S}\left\{\left(1+\frac{2-\eta}{\delta}\right)+(2-\eta)\sum_{i=a,b,c}\left(\frac{1}{2}+\frac{1}{\delta}\right)\left(\frac{2T_{c}\alpha_{i}\sigma_{i}^{1/\delta}}{T_{h}\alpha_{h}\sigma_{h}^{1/\delta}}(2-\eta)\right)^{\frac{\delta}{\delta+1}}\right\}\right]^{\delta},
\label{thnp}
\end{equation}

\begin{equation}
\tilde{t}_{a}=\left[\frac{T_{c}\alpha_{a}\sigma_{a}^{1/\delta}}{(T_{h}-T_{c})\Delta S}\left\{\left(1+\frac{2}{\delta}\right)\left(1+\sum_{i=b,c}\left(\frac{\alpha_{i}\sigma_{i}^{1/\delta}}{\alpha_{a}\sigma_{a}^{1/\delta}}\right)^{\frac{\delta}{\delta+1}}\right)+\left(\frac{2}{2-\eta}+\frac{2}{\delta}\right)\left(\frac{T_{h}\alpha_{h}\sigma_{h}^{1/\delta}}{2T_{c}\alpha_{a}\sigma_{a}^{1/\delta}}(2-\eta)\right)^{\frac{\delta}{\delta+1}}\right\}\right]^{\delta},
\label{tanp1}
\end{equation}

\begin{equation}
\tilde{t}_{b}=\left[\frac{T_{c}\alpha_{b}\sigma_{b}^{1/\delta}}{(T_{h}-T_{c})\Delta S}\left\{\left(1+\frac{2}{\delta}\right)\left(1+\sum_{i=a,c}\left(\frac{\alpha_{i}\sigma_{i}^{1/\delta}}{\alpha_{b}\sigma_{b}^{1/\delta}}\right)^{\frac{\delta}{\delta+1}}\right)+\left(\frac{2}{2-\eta}+\frac{2}{\delta}\right)\left(\frac{T_{h}\alpha_{h}\sigma_{h}^{1/\delta}}{2T_{c}\alpha_{b}\sigma_{b}^{1/\delta}}(2-\eta)\right)^{\frac{\delta}{\delta+1}}\right\}\right]^{\delta},
\label{tbnp2}
\end{equation}
and 
\begin{equation}
\tilde{t}_{c}=\left[\frac{T_{c}\alpha_{c}\sigma_{c}^{1/\delta}}{(T_{h}-T_{c})\Delta S}\left\{\left(1+\frac{2}{\delta}\right)\left(1+\sum_{i=a,b}\left(\frac{\alpha_{i}\sigma_{i}^{1/\delta}}{\alpha_{c}\sigma_{c}^{1/\delta}}\right)^{\frac{\delta}{\delta+1}}\right)+\left(\frac{2}{2-\eta}+\frac{2}{\delta}\right)\left(\frac{T_{h}\alpha_{h}\sigma_{h}^{1/\delta}}{2T_{c}\alpha_{c}\sigma_{c}^{1/\delta}}(2-\eta)\right)^{\frac{\delta}{\delta+1}}\right\}\right]^{\delta}.
\label{tbnp3}
\end{equation}
\end{widetext}

The ratios of $\frac{\tilde{t}_{h}}{\tilde{t}_{i}} (i : a, b, c)$ can also be obtained from the optimized efficient power $\chi_{ep}$ and are given below:
\begin{equation}
\left(\frac{\tilde{t}_{h}}{\tilde{t}_{i}}\right)^{\left(\frac{1}{\delta}+1\right)} = \frac{T_{h}\alpha_{h}\sigma_{h}^{1/\delta}}{2T_{c}\alpha_{i}\sigma_{i}^{1/\delta}}(2-\eta).
\label{ratio1}
\end{equation}
Similarly the ratios for $\frac{\tilde{t}_{i}}{\tilde{t}_{j}}$  with $i, j = a, b, c$ are also given by,
\begin{equation}
\left(\frac{\tilde{t}_{i}}{\tilde{t}_{j}}\right)^{\left(\frac{1}{\delta}+1\right)} = \frac{\alpha_{i}\sigma_{i}^{1/\delta}}{\alpha_{j}\sigma_{j}^{1/\delta}}.
\label{ratio2}
\end{equation}
Similar to that done for $\dot{\Omega}$ figure of merit, Eq.(\ref{eta}) on further substitution of Eq.(\ref{thnp}) and Eq.(\ref{ratio1}) with $\eta =\eta_{\chi_{ep}}$  yields the efficiency at maximum efficient power, $\chi_{ep}$ and is given by, 
\begin{equation}
\eta_{\chi_{ep}} =\frac{\eta_{C}}{\delta}\left\{\frac{\phi+\phi^{\frac{1}{\delta+1}}\xi}{1+\frac{\phi}{\delta}+\phi^{\frac{1}{\delta+1}}\xi\left(\frac{1}{2}+\frac{1}{\delta}\right)-\eta_{C}}\right\}
\label{npfinal}
\end{equation}
where in the above equation, $\phi=2-\eta_{\chi_{ep}}$ and $\xi = \left(\frac{2T_{c}\alpha_{i}\sigma_{i}^{1/\delta}}{T_{h}\alpha_{h}\sigma_{h}^{1/\delta}}\right)^{\frac{\delta}{\delta+1}}$. It is to be noted that the above expression contains $\eta_{\chi_{ep}}$ on both sides which is too complicated to solve in the generalized fashion.  However, the solution with $\delta=1$ (low dissipation) is found when neglecting the adiabatic dissipation co-efficients, $\sigma_{a}=0$ and $\sigma_{b}=0$, which  is same as the efficiency derived for Carnot like heat engines without adiabatic dissipation \cite{singh}. 
Similar to the efficiency at maximum $\dot{\Omega}$ figure of merit, the value of efficiency at maximum efficient power $\eta_{\chi_{ep}}$ also depends on the ratio between values of $\sigma_{i}$'s and $\sigma_{h}$. The generalized extreme bounds on the efficiency at maximum $"\chi_{ep}"$ are obtained when $\sigma_{h}\rightarrow 0$ and $\sigma_{h}\rightarrow \infty$. That is, when $\sigma_{h}\rightarrow 0$, 
$\xi\rightarrow\infty$ which gives $\eta_{\chi_{ep}}=\frac{2\eta_{C}}{\delta+2}$ and when $\sigma_{h}\rightarrow \infty$, 
$\xi\rightarrow 0$ which provides $\eta_{\chi_{ep}}=\frac{\Upsilon+\sqrt{\Upsilon^{2}-8\eta_{C}}}{2}$, where $\Upsilon = (\delta+2)-\eta_{C}(\delta-1)$.  This shows that the efficiency at the maximum efficient power lies between these two extreme bounds, which is given by,
\begin{widetext}
\begin{equation}
\frac{2\eta_{C}}{\delta+2} \equiv\eta^{-}_{\chi_{ep}}\leq\eta_{\chi_{ep}}\leq\eta^{+}_{\chi_{ep}}\equiv
\frac{(\delta+2)-\eta_{C}(\delta-1)+\sqrt{((\delta+2)-\eta_{C}(\delta-1))^{2}-8\eta_{C}}}{2}.
\label{etaetalimit}
\end{equation}
\end{widetext}

Thus the generalized lower and upper bounds of efficiency at maximum efficient power are obtained for the asymmetric dissipation limits of  $\sigma_{h}\rightarrow 0$ and $\sigma_{h}\rightarrow \infty$ respectively, for any finite values of $\sigma_{i} (i : a, b, c)$.
When $\delta = 1$, the values of optimized efficiency at the maximum efficient power at low dissipation regime is obtained, which is $\frac{2\eta_{C}}{3}$ and $\frac{3+\sqrt{9-8\eta_{C}}}{2}$, the lower and upper bound respectively \cite{holubec, singh}. 
Similar to $\dot{\Omega}$ figure of merit, this result also shows that the inclusion of finite time non-adiabatic dissipation on low-dissipation model does not influence the lower and upper bound on the efficiency optimized at maximum efficient power.  
It is to be noted from Eq.(\ref {npfinal}) that when $\delta \to 0$ (no dissipation), $\eta^{-}_{\chi_{ep}}$ and $\eta^{+}_{\chi_{ep}} \to \eta_C$ and when $\delta \to \infty$ (high super dissipation limit), $\eta^{-}_{\chi_{ep}}$ and $\eta^{+}_{\chi_{ep}} \to 0$. 
 Thus, the generalized universal nature of lower and upper bounds on the efficiency of Carnot like heat engines at maximum efficient power, $\eta_{\chi_{ep}}$ (Eq. \ref{etaetalimit}) under the combinations of isothermal and adiabatic asymmetric dissipation limits is obtained.

\section{Conclusion}
In this paper, the generalized extreme bounds of the efficiency for the power law dissipative Carnot-like heat engines under $\dot{\Omega}$ figure of merit and efficient power $\chi_{ep}$ optimization criteria was investigated. Since, the $\dot{\Omega}$ figure of merit  provides the trade off between the useful energy delivered and energy lost of heat engine and the efficient power governs the compromise between power and efficiency of heat engine,  finding generalized bounds of the efficiency with these target functions are very relevant in direct correlation of the actual needs of energy consumption, availability of resources and environmental impact. Too much mathematical complexity is observed while obtaining the generalized expression for optimized efficiency at maximum $\chi_{ep}$ as compared to $\dot{\Omega}$ figure of merit.  When $\delta= 1$, the bounds of the efficiency with the $\dot{\Omega}$ figure of merit and efficient power $\chi_{ep}$ in the asymmetric dissipation converges to the same bounds as the corresponding ones obtained from previous low dissipation model. In corroborated with the efficiency at maximum power, these results also showed that the presence of non-adiabatic dissipation does not influence the minimum and maximum bounds on the efficiency optimized at maximum $\dot{\Omega}$ figure of merit and also maximum efficient power obtained in the low dissipation model which does not take in to account the non-adiabatic dissipation. The future work will focus on comparison of these figure of merit predictions with different heat engine models and with observed efficiency of real heat engines \cite{medinasl}.

\end{document}